Plasmonically-Enhanced Emission from an Inverted GaN Light Emitting Diode

Michael A. Mastro[a], Byung-Jae Kim [b], J. A. Freitas, Jr.[a], Joshua D. Caldwell[a], Ron Rendell[a], Jennifer Hite[a], Charles R. Eddy, Jr.[a], Jihyun Kim[b]

[a] US Naval Research Laboratory
[b] Department of Chemical and Biological Engineering, Korea University

**Abstract**
Silver nanoparticles dispersed on the surface of an inverted GaN LED were found to plasmonically enhance the near-bandedge emission. The resonant surface plasmon coupling led to a significant enhancement in the exciton decay rate and the ensemble of nanoparticles provided a mechanism to scatter the coupled energy as free space radiation. The inverted LED structure employed a tunnel junction to avoid the standard thick p+ GaN current spreading contact layer. In contrast to a standard design, the top contact was a thin n++ AlGaN layer, which brought the quantum well into the fringing field of the silver nanoparticles. This proximity allowed the excitons induced within the quantum well to couple to the surface plasmons, which in turn led to the enhanced band edge emission from the LED.

**INTRODUCTION**

Electron-hole pairs generated in the active region of a light emitting diode decay radiatively by emitting a photon or non-radiatively as dissipated energy in the crystal lattice. Modification of the surrounding density of states can increase or decrease the spontaneous emission process as described by Fermi's golden rule. Introduction of a plasmonic metal in the near vicinity (tens of nanometers) of the semiconductor active region introduces an alternate decay route. This coupling between the excitons within the active region and the surface plasmons within the plasmonic metal occurs when the energy of the electron-hole pair has sufficient overlap in energy with the energy of the surface plasmon resonance condition. Okamoto et al. found large enhancements in the emission from InGaN quantum well (QW) devices when a thin film of platinum or silver was deposited 10 nm above QW.[1] At a metal/dielectric interface the surface plasmon fields exist as evanescent waves that decay exponentially into the nearby semiconductor, thus the close proximity of the QW to the noble metal is required for the QW excitons to be located within the fringing field of the surface plasmons. Okamoto et al. calculated the fringing field depth as 47 and 77 nm for silver and aluminum, respectively.

The surface plasmon at a perfectly flat metal/semiconductor interface is a non-propagating evanescent wave. Without scattering, this energy will be non-radiatively absorbed and dissipated as heat.[2] Okamoto et al. relied on imperfections and roughness in the metal coating to couple the surface plasmons into scattered light.[1] Neogi and Morkoc deposited arrays of silver nanoparticles onto a structure with GaN quantum dots located in close proximity to the surface.[3] Structuring the silver as nanoparticles allows for tuning of the surface plasmon resonance condition where the surface plasmon DOS is extremely high, to closely match the peak emission of the active layer of the semiconductor emitter. Furthermore, the periodicity of the silver nanoparticles allows a more controlled scattering of the surface plasmon into a radiative photon.

The necessity of placing the active region of the semiconductor device within approximately 30 nm of the plasmonic metal is a subtle dilemma for (In,Al)GaN based LEDs. Kwon et al. avoided this issue by placing the silver nanoparticles internal to the semiconductor; specifically, the plasmonic metal layer was situated in the n-GaN within a few nanometers of the multi-QW.[4] However, any foreign particle located within the crystalline structure of the semiconductor should disturb the crystallographic stacking order during epitaxy, although Kwon et al. maintained that the crystal was not perturbed by the discontinuous silver interlayer. Surface plasmons of the silver interlayer that do scatter are emitted as photons within the semiconductor slab, which will suffer from extraction issues (e.g., total internal reflection) that are problematic for all LEDs.

A major vein of LED research is enhancing the extraction of light from the semiconductor into the surrounding air by employing a microcavity, photonic crystal, or surface roughening.[5-8] Another advantage of enhancing spontaneous emission via plasmonic metal particles located exterior to the semiconductor slab is that energy couples from the active region to the surface plasmons within the metal, thus avoiding the necessity for light to exit via the semiconductor/air light cone.

The final grown layer in a traditional GaN-based LED structure is a p-type GaN contact layer to avoid magnesium memory effects and doping generated growth defects. The magnesium dopant has a high activation energy, which limits the experimentally achievable density of active acceptors to approximately $10^{18}$ cm$^{-3}$ at room temperature. For a simple pn junction, this corresponds to a p side depletion width of 32.2nm and an n side (with $5 \times 10^{18}$cm$^{-3}$ active donors) depletion width of 6.5nm. Once depletion from the surface is included the necessary theoretical thickness of the p-type region is already beyond the surface plasmon fringing field thickness, making exterior surface plasmon enhancements impractical for such structures. In practice, the high level of defects also results in low carrier mobility and thus high resistivity of the p-GaN, which necessitates a thick (100 to 250 nm) current-spreading p-type layer in commercial LEDs, which only exacerbates this issue.

Low resistance Ohmic contacts to the p-type region require a high level of magnesium doping. To avoid the growth of and issues related to contacting the p$^+$ GaN contact layer, an inverted LED structure has been suggested by commercial entities.[9,10] Operation is similar to a standard LED in that electrons and holes from the n- and p-type regions, respectively, are injected into a QW active region; however, this structure is advantageous as both Ohmic contacts are made to n-type material. This creates an npn semiconductor that would prohibit current flow in either the forward or reverse direction for non-degenerate doping levels. The key to an inverted LED biased by two n-type contacts is the formation of a (degenerate) tunnel junction that under local reverse bias allows electrons to quantum-mechanically tunnel through a thin depletion region and then enter the active region of the device where they may recombine radiatively.

This article reports on the direct observation of plasmonic coupling via a semi-ordered array of silver nanoparticles on an inverted LED. The inverted LED structure has only a thin n$^+$ AlGaN contact layer above the active region, which allows surface plasmon coupling to the excitons generated in the QW.

**Experimental**
The inverted LED structures as depicted in Fig. 1 were grown in an impinging-flow metal-organic chemical vapor deposition system. The structure in Fig 1a was grown in two steps to avoid the Mg memory effect. Specifically a set of 2μm p-type layers were grown on double sided polished sapphire in a separate set of experiments. The reaction chamber was cleaned, and subsequently the active and n-type contact layer were deposited on the p-type template.

The tunnel junction structure displayed in Fig 1b avoids the necessity for a thick p-type layer, and the structure could be grown in one continuous step. The p- and n-type doping in the tunnel junction were sufficiently high to eliminate the p- and n- depletion barriers, thereby enabling the electrons to pass directly through the junction. This in turn allowed a non-rectified current flow from the 4H-SiC into the n+

GaN cap layer and thus successful operation of the LED.

A discontinuous silver film with a target thickness of approximately 100 nm was deposited in an e-beam evaporator onto the surface of the LED structure. The silver islands were exposed to a rapid thermal anneal for 1 min at 700°C. This resulted in an ensemble of silver nanoparticles dispersed on the sample surface with an average diameter of 150 nm (Fig. 2).

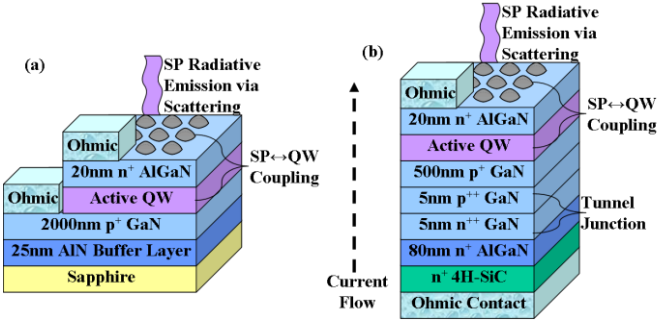

Figure 1. Schematic of an inverted LED (a) with and (b) without a tunnel junction. The thin n+ AlGaN contact layer separates the 5nm AlGaN / 4nm GaN / 5nm AlGaN quantum well active region from the distribution of silver nanoparticles on the surface.

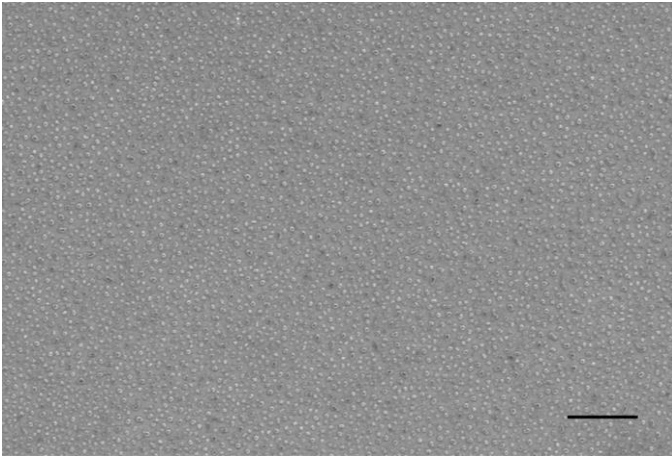

Figure 2. Scanning electron micrograph of silver nanoparticles on the surface of the inverted LED structure. Scale bar is 1000nm.

**Results**

A two-dimensional coupled dipole approximation method was employed to model the electromagnetic behavior of an array of silver nanoparticles on the surface of AlGaN surrounded by air.[11,12] Fig. 3 shows that alloying AlN into the GaN contact layer leads to a blue-shift in the extinction peak. In this work the alloy $Al_{0.1}Ga_{0.9}N$ composition was selected to improve injection efficiency into the GaN quantum well. Fig. 3 shows that a peak in extinction efficiency was found at 365 nm for the array of silver nanoparticles with a diameter of 150 nm, and particle to particle center spacing of 225 nm situated at an $Al_{0.1}Ga_{0.9}N$ / air interface. The rapid thermal annealing conditions were selected to produce an set silver nanoparticles that closely resembled this geometry, and thus match the surface plasmon resonance to the quantum well near bandedge emission.

In addition, the model showed a complex dependence on inter-particle coupling relative to the single particle Mie resonance. Confinement of electrons in small noble-metal nanoparticles leads to an electromagnetic resonance upon plane-wave excitation. The charge within the silver nanoparticles move in phase, which creates an effective restoring force at the particle dipole plasmon frequency.[12] Thus, leading to a corresponding dipolar field forming in the near vicinity of the nanoparticles. Still, the electron-hole pairs in the metal can couple to the phonon bath, which leads to Ohmic heating. In essence the imaginary component of the metal dielectric response limits the polarizability.

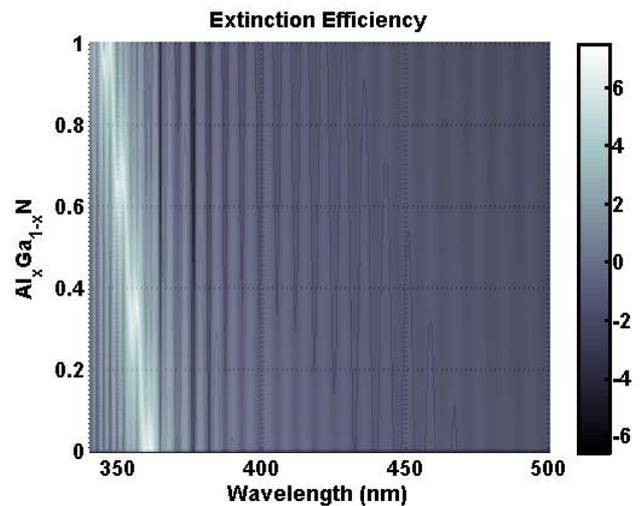

Figure 3. Calculation of extinction efficiency via a coupled dipole interaction model for array of silver nanoparticles situated at an $Al_xGa_{1-x}N$ / air interface. The peak resonance is visible as the bright stripe near approximately 350nm. The decrease in dielectric constant for increasing AlN content in the AlGaN film shifts the peak in extinction efficiency further into the UV.

The luminescence of the sample was excited at 5K with the 325-nm line of a He–Cd laser is displayed in Fig 4. The uncorrected photoluminescence spectra showed a factor of 5.2 and 3.6 increase (on two separate samples) in near bandedge emission from the silver coated structure in comparison to the emission prior to deposition. The wavelength of this enhancement (~365 nm) coincided with the surface plasmon resonance condition.

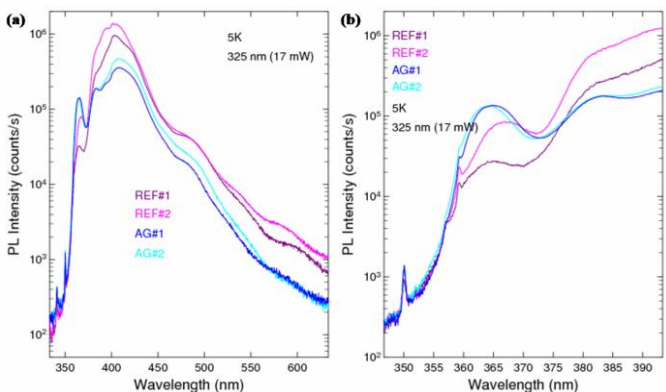

Figure 4. Photoluminescence acquired at two different spectral ranges of an inverted LED with a silver nanoparticle coating. Two characteristic sample sets (Ref#1, Ag#1 and Ref#2,Ag#2) are displayed.  The silver nanoparticles were designed to have a plasmon resonance coinciding with the 365nm near bandedge emission of the GaN quantum well. The coupling with the on-resonant 365nm emission led to an average approximate four times increase in luminescence.

A time-resolved photoluminescence measurement found that the addition of the silver nanoparticles increased the photoluminescence decay rate by a factor of 3.81. Excitons generated in the active region can decay radiatively as a photon, non-radiatively as dissipated heat in the lattice, or the energy can be transferred to the surface plasmons. At this point, the coupled surface plasmon energy in the metal can decay as dissipated heat, which in turn would quench the emission. The increase in decay rate indicates that an alternate mechanism is provided by the nanostructure of the particles, which creates an ensemble of scattering sites that bridges the momentum mismatch to rapidly convert the energy to free space radiation.[13] The similar increase in the intensity and the decay ratio of the photoluminescence confirms that the spontaneous emission wavelength-dependent enhancement was clearly strongest at the surface plasmon resonance energy where the surface plasmon DOS is high.[13,14]

**Summary**

Silver thin films and nanoparticles deposited on the surface of InGaN quantum well structures have been shown by several groups to plasmonically enhance the optically pumped emission. The necessity of placing the active region of the device within approximately 30 nm of the plasmonic metal is a subtle dilemma for GaN-based LEDs. Hence, nearly all demonstrations of exciton to surface plasmon coupling have been based on quantum well structures that are not designed for electroluminescence.

In this paper an approach was described for plasmonic enhancement of an inverted LED structure. A variety of inverted LED structures were developed with the key design of a thin n++ (Al)GaN top contact, which brings the quantum well into the fringing field of the silver nanoparticles. This proximity allowed the excitons induced within the quantum well to couple to the surface plasmons, thereby increasing the emission by approximately a factor of four in a functional LED. Furthermore, the inverted LED allows flexibility in the alloy composition of the top contact layer, which alters the dielectric environment and plasmonic response of the metal.  The approach of this paper can be extended to any inverted (Al,In)GaN LED by tailoring the size of the noble metal nanoparticles and consequently the location of the resonant surface plasmon energy to match the UV, violet, blue, or green emission.

**Acknowledgements**

Research at NRL was supported by the ONR. The work at Korea University was supported by the Carbon Dioxide Reduction and Sequestration Center, one of the 21st Century Frontier R&D Program funded by the Ministry of Education, Science and Technology of Korea.

**References**

1. K. Okamoto, I. Niki, A. Shvartser, Y. Narukawa, T. Mukai, and A. Scherer, Nature Materials, Vol. 3, No. 9, pp. 601-605 (2004).
2. C. Langhammer, Z. Yuan, I. Zoric, B. Kasemo, Nano Lett. 6, 833 (2006)
3. Neogi, H. Morkoc, Nanotech. 15, 1252 (2004)
4. M.-K. Kwon, J.-Y. Kim, B.-H. Kim, I.-K. Park, C.-Y. Cho, C.C. Byeon, S.J. Park, Adv. Mat. 20, 1253 (2008)
5. M.A. Mastro, J.D. Caldwell, R.T. Holm, R.L. Henry, C.R. Eddy Jr., Adv. Mater., 20-1, 115 (2008)
6. M.A. Mastro, C.S. Kim, M. Kim, J. Caldwell, R.T. Holm, I. Vurgaftman, J. Kim, C.R. Eddy Jr., J.R. Meyer, Jap. J. Appl. Phys. 47, 7827 (2008)
7. B.-J. Kim, M.A. Mastro, H. Jung, H.-Y. Kim, S. H. Kim, R.T. Holm, J. Hite, C.R. Eddy Jr., J. Bang, J. Kim, , Thin Solid Films, 516-21, 7744 (2008)
8. B.-J. Kim, H. Jung, J. Shin, M.A. Mastro, C.R. Eddy Jr., J.K. Hite, S.H. Kim, J. Bang, J. Kim, Thin Solid Films, 517-8, 2742 (2009)
9. J.A. Edmond, K.M. Doverspike, M.J. Bergmann, H.-S. Kong, Inverted Light Emitting Diode on Conducting Substrate, US Patent Application No. 10/367,495 (2004)
10. T. Takeuchi, G. Hasnain, S. Corzine, M. Hueschen, R.P. Schneider, Jr., C. Kocot, M. Blomqvist, Y.-l. Chang, D. Lefforge, M.R. Krames, L.W. Cook, S.A. Stockman, Jpn. J. Appl. Phys. 40, L861 (2001)
11. L. Zhao, L. Kelly, G.C. Schatz, J. Phys. Chem. B 107, 7343 (2003)
12. R. Kullok, S. Gradstrom, P/ Evans. R. Pollard, L. Eng, J. Opt. Soc. Am. B. 27, 1819 (2010)
13. K. W. Liu, Y. D. Tang, C. X. Cong, T. C. Sum, A. C. H. Huan, Z. X. Shen, Li Wang, F. Y. Jiang, X. W. Sun and H. D. Sun, Appl. Phys. Lett., 94, 151102 (2009)
14. M.A. Mastro, J. A. Freitas, Jr., O. Glembocki, C.R. Eddy, Jr., R. T. Holm, R.L. Henry, J. Caldwell, R.W. Rendell, F. Kub, J-H. Kim, Nanotech. 18, 265401 (2007)